# Revealing time characteristics of optical excitations in dielectric and plasmonic structures through cathodoluminescence interferometry

Evelijn Akerboom[a*], Hiroshi Sugimoto[b], Minoru Fujii[b], Nicolas Pazos-Perez[c], Ramon A. Álvarez Puebla[c,d], A. Femius Koenderink[a], F. Javier García de Abajo[d,e], and Albert Polman[a*]

[a]Center for Nanophotonics, NWO-Institute AMOLF, Science Park 104, 1098 XG Amsterdam, the Netherlands

[b] Department of Electrical and Electronic Engineering, Graduate School of Engineering, Kobe University, Rokkodai, Nada, Kobe 657-8501, Japan

[c] Department of Physical and Inorganic Chemistry, Universitat Rovira i Virgili, 43007, Tarragona, Spain

[d] ICREA-Institució Catalana de Recerca i Estudis Avançats, Passeig Lluís Companys 23, 08010 Barcelona, Spain

[e] ICFO-Institut de Ciencies Fotoniques, The Barcelona Institute of Science and Technology, 08860 Castelldefels, Barcelona, Spain

* Corresponding email: e.akerboom@amolf.nl, a.polman@amolf.nl

---

## Abstract

Cathodoluminescence (CL) spectroscopy provides access to optical excitations with nanometer spatial resolution, but direct time-resolved measurements of optical resonances remain challenging. Here, we demonstrate that CL interferometry provides access to the temporal response, phase behavior, and modal spectral structure of resonant nanoscale scatterers without requiring ultrafast pump-probe schemes. We develop an analytical framework in which Fourier transformation angle- and frequency-resolved CL interferograms yields the decay time of optical resonances governed by the linear optical response. Multimode resonators exhibit characteristic temporal CL beating signatures associated with spectral mode splitting. By exploiting transition radiation emitted from a nearby metallic surface as a broadband reference, we further demonstrate phase retrieval and cross-correlation measurements between instantaneous and resonant emission processes. Experimental measurements on Au nanoparticles, broadband plasmonic emitters, Au nanostars, and Si nanospheres supporting multipolar Mie resonances confirm the theoretical predictions, and decay times in the range 1-10 fs are derived for each system. Our results establish CL interferometry as a powerful approach for accessing spectral, spatial, and phase information within a single nanoscale measurement with fs resolution.

The temporal characteristics of optical resonances play a central role in nanoscale light-matter interactions, governing how electromagnetic energy is stored, redistributed, and dissipated in nanophotonic systems. Resonance lifetimes and phase behavior determine the efficiency of processes ranging from plasmon-enhanced spectroscopy and photocatalysis to nanoscale heat generation and energy conversion[1–4]. In nanostructures, optical excitation of localized surface plasmon resonances (LSPRs) or Mie resonances is followed by ultrafast dephasing and relaxation processes occurring on femtosecond timescales[5–7]. Owing to these extremely short lifetimes, direct measurements of resonance dynamics typically require sophisticated ultrafast techniques such as pump-probe spectroscopy [8–10]. While such methods provide excellent temporal resolution, combining femtosecond sensitivity with nanometer-scale spatial resolution remains a major challenge.

Cathodoluminescence (CL) spectroscopy has emerged as a powerful approach to probe optical excitations with nanoscale spatial resolution, using fast electrons as broadband sources of optical excitations. CL spectroscopy has been used to map optical properties of photonic nanostructures both spectrally and spatially[11–14], and has recently been extended to include time-resolved measurements[15–17]. However, due to the limited temporal resolution of CL pump-probe schemes, direct experimental access to resonance phase and dephasing dynamics that occur at the femtosecond time scale remains challenging.

Recent developments in CL interferometry have demonstrated that spectral interference between multiple optical pathways can be used to recover temporal information from CL measurements[18]. Since the linear temporal response of an optical system is related to its spectral response through Fourier-transform relations[19], interferometric measurements provide a route to access resonance temporal response and phase information without requiring ultrafast laser excitation[20–23]. Here, we experimentally demonstrate that CL interferometry provides direct access to the temporal dynamics of excited resonant nanoscale scatterers on femtosecond timescales. We develop an analytical framework relating the Fourier transformation of CL interferograms to resonance dephasing time, phase, and modal structure, and validate these concepts experimentally using plasmonic and dielectric nanostructures.

We study the CL excitation of nanoparticles that are placed above a reflecting substrate, resulting in angular and spectral interferograms due to the interference of the direct and the substrate-scattered CL emission. Superimposed on this is the coherent transition radiation that is excited by the electron upon impact on the substrate. The electric far-field amplitude at frequency $\omega$ and in-plane (i.e., parallel to the substrate) wave vector $\mathbf{k}_\parallel$ can then be written as

$$\mathbf{E}(\mathbf{k}_\parallel, \omega) = P(\omega)A(\omega)\mathbf{S}(\mathbf{k}_\parallel, \omega)\left[e^{ik_z h} + r(\mathbf{k}_\parallel, \omega)e^{-ik_z h}\right] + \mathbf{E}_{TR}(\omega, k_\parallel)e^{i\phi_0}, \qquad (1)$$

where $P(\omega)$ is a characteristic of the driving pulse (the electron), $A(\omega)$ is the particle's scattering amplitude, $\mathbf{S}(\mathbf{k}_\parallel, \omega)$ is a spherical wave vector amplitude that depends on the scattering resonance type (dipole, quadrupole, etc.), $k_z = k\cos\theta$ is the longitudinal wave vector (along the substrate normal, parallel to the electron beam (e-beam)), with $\theta$ the azimuthal angle defined with respect to the substrate surface, $h$ is the distance between the particle and the substrate, and $r(\mathbf{k}_\parallel, \omega)$, is the reflection coefficient of the latter.

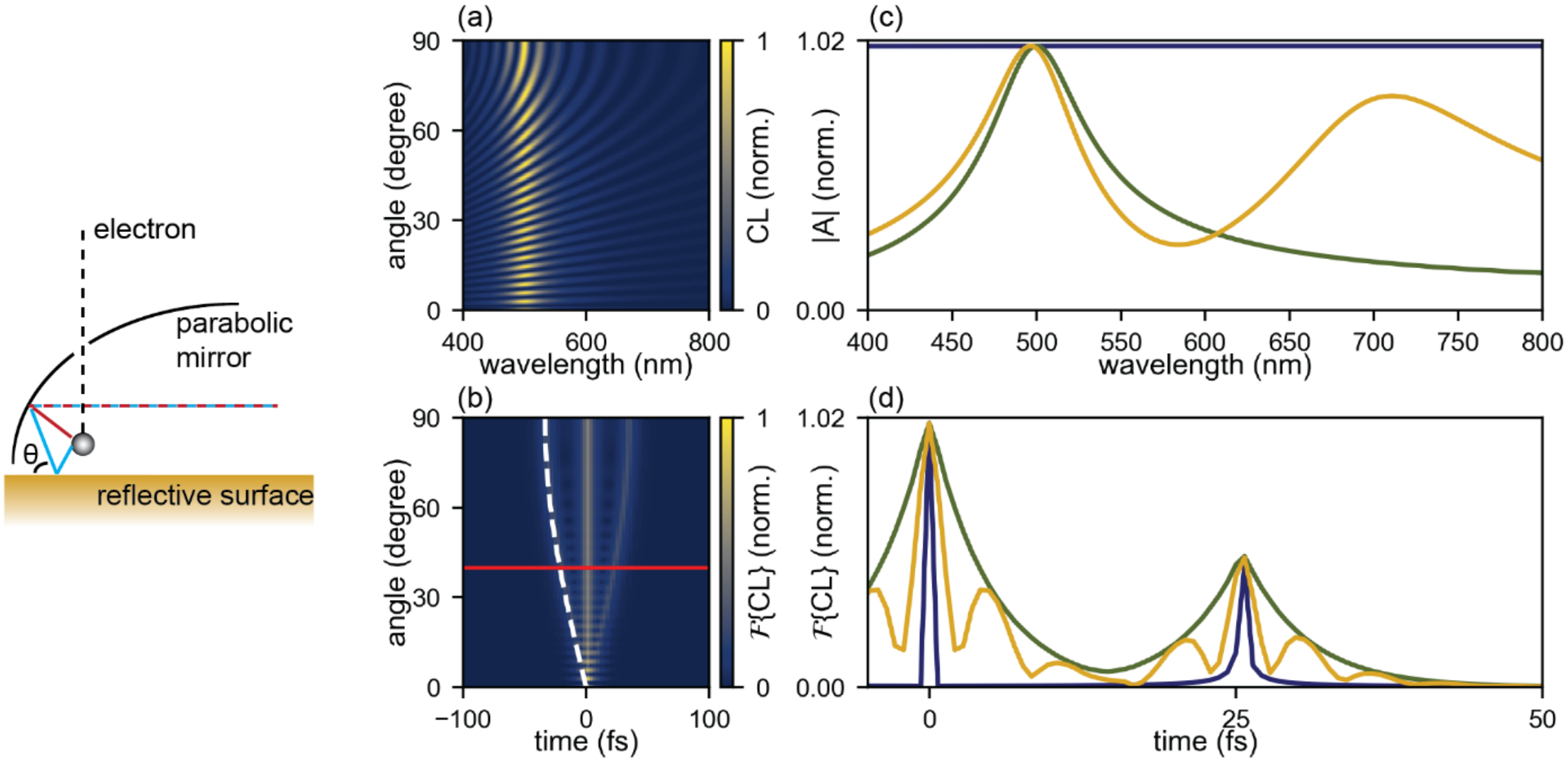

*Figure 1. Analytical calculation of CL interferometry of a radiating resonant nanoparticle placed at a distance h = 5 μm above a reflecting surface. (a) Angle- and wavelength- resolved CL interferogram of a Lorentzian scatterer with a resonance wavelength of 500 nm and a linewidth of 0.2 rad/fs, and (b) its Fourier transform to the time domain. The dashed curve in (b) shows the calculated time delay associated with the interference pattern for each emission angle. (c) Optical response function of three distinct resonances: a broadband emitter (blue constant line at 1.0), the Lorentzian scatterer from (a,b) (green), and a scatterer with two scattering resonances, one at $\lambda$=500 nm with $\gamma$=0.2 rad/fs and another at $\lambda$=700 nm with $\gamma$=0.25 rad/fs (yellow). (d) Line scans of the Fourier-transform of the calculated CL interferogram at an emission angle of $40°$ for the three spectral response functions in (c).*

The rightmost term in Eq. (1) ($\mathbf{E}_{TR}$) gives the contribution from transition radiation (TR) excited at the reflecting surface, using a phase delay $\phi_0 = \omega h / v_e$, with $v_e$ the speed of the electron, to take into account the time of flight of the electron to travel from the particle to the surface. We assume the particle is a Lorentzian scatterer of resonance frequency $\omega_0$ and linewidth ($\gamma_0$) given by

$$A(\omega) = \frac{A_0}{\omega_0 - \omega - i\gamma_0}. \tag{2}$$

When multiple scattering resonances from the same particle contribute, the corresponding terms can be added to the scattering and spherical wave amplitude functions. For simplicity, we assume all resonances to have a Lambertian angular emission profile. The total CL intensity in the interferogram is given by

$$I(\mathbf{k}_{\parallel}, \omega) = \left|E(\mathbf{k}_{\parallel}, \omega)\right|^2. \tag{3}$$

We use the formalism above to study the effect of the scattering spectrum and linewidth on the interferograms. We first study a configuration without TR. We use a scatterer placed at a distance *h* = 5 μm with a resonance wavelength at 500 nm and a linewidth of 0.2 rad/fs, corresponding to a dephasing time of 4.4 fs. For simplicity, we first assume a perfectly reflecting substrate ($r = 1$). We observe an interference pattern with fringes whose period decreases for higher emission angle due to the increased path length difference for the interfering signals (Figure 1(a)). This is also clear in Fig. 1(b), which shows the Fourier transform of Fig. 1(a) from frequency to time. It shows that the calculated time delay associated with the interference pattern for each angle is given by $\Delta t = 2h\sin(\theta)/c$, which overlays the result rather well.

Next, we study the linewidths in the Fourier transforms for particles with three different optical responses as shown in Fig. 1(c): a broadband scatterer (blue constant line), a Lorentzian scatterer as was used in Fig. 1(a,b) (green), and a scatterer with two scattering resonances, one at $\lambda=500$ nm with $\gamma=0.2$ rad/fs and another at $\lambda=700$ nm with $\gamma=0.25$ rad/fs (yellow). The corresponding Fourier transforms at an emission angle of $40°$ (Fig. 1(d)) show some distinct differences. First, all time traces show a peak at a time delay of 26 fs, associated with the path length difference of the interfering signals[23]. The linewidth differs substantially for different peaks: it is small for the broadband scatterer due to its instantaneous response, while for the resonant scatterer, the linewidth represents the decay time of the resonance. For the scatterer with two resonances, we observe an envelope that follows the green line, with underlying oscillations that represent the beating of the two resonances with a difference of the resonance periods of 5.8 fs. To summarize, we have shown that angle- and spectrally resolved CL data in this system can be seen as an analog to conventional spectral pulse interferometry: in this analogy the reflecting substrate creates a copy of the pulse emitted by the nanoparticles, and the viewing angle gives a method to observe pulse interferograms as function of pulse delay. The time-resolved features are thus similar to autocorrelations of the pulsed sample emission.

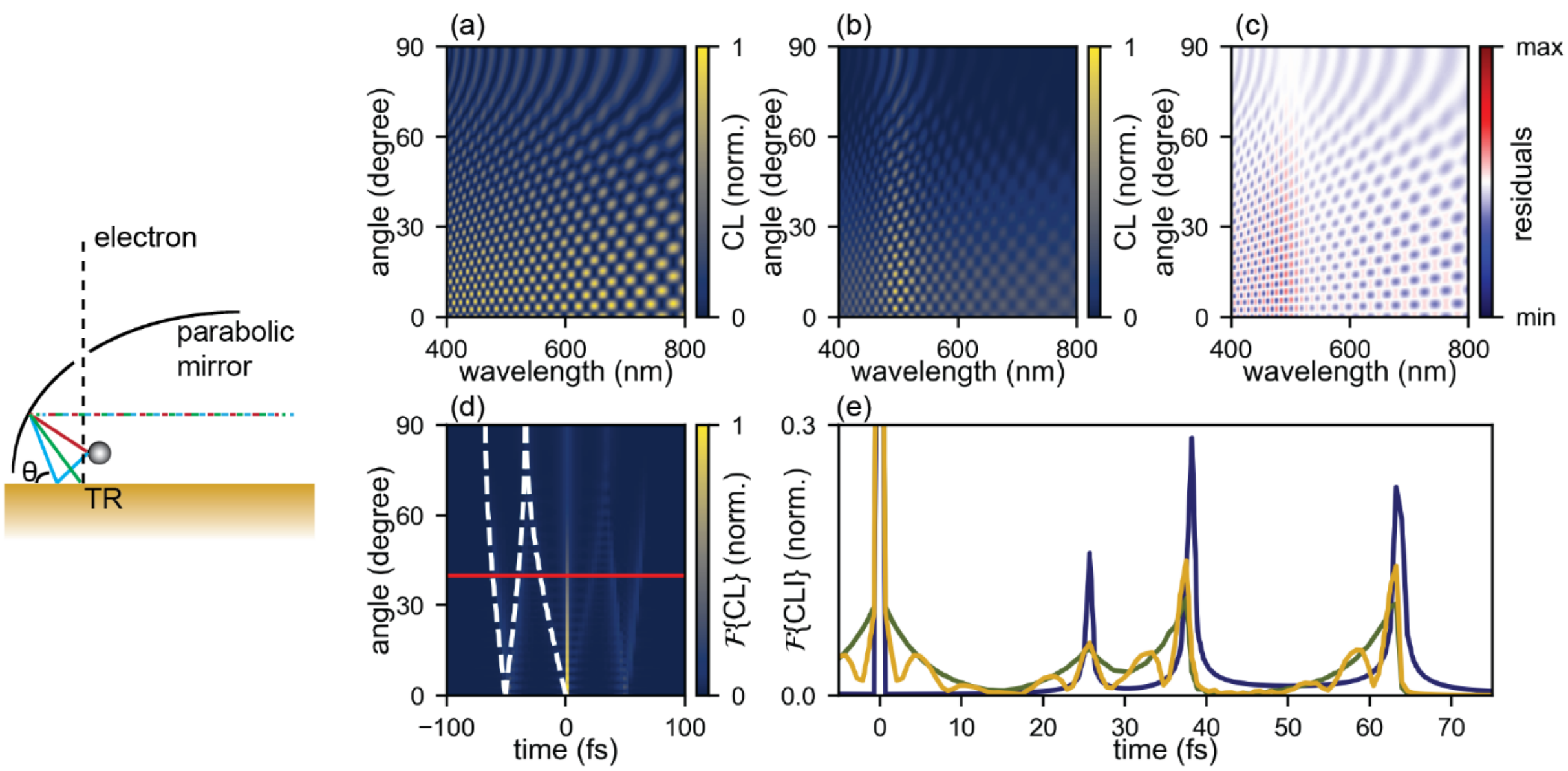

*Figure 2. Analytical calculation of CL interferometry of a resonantly scattering particle placed above a reflecting surface, including TR emission from the substrate. Angle- and wavelength- resolved CL interferogram for (a) a broadband optical emitter, and (b) a Lorentzian scatterer with a resonance wavelength of 500 nm and a linewidth of 0.2 rad/fs. (c) Residuals of (a) and (b). (d) Fourier transform of (b) to the time domain, where the dashed curves show the calculated time delays for each angle, and (e) line scans of the Fourier-transform of the calculated CL interferograms at an emission angle of $40°$ for the three spectral response functions in Fig. 1(c).*

Next, we study a configuration in which a 30 keV electron first excites the nanoparticle and then impinges on a substrate and generates TR. Figures 2(a) and 2(b) show the interference patterns for a broadband scatterer and a resonant scatterer ($\lambda=500$nm with $\gamma=0.2$ rad/fs), respectively. In both of them, we observe a rich interference pattern associated with the additional interference terms originating from the coherent contribution of TR. This is clear in Fig. 2(d), where the Fourier transform of Fig. 2(b) shows two branches at longer delay time, determined by the time of flight of the electron from the nanoparticle to the substrate[18].

Fourier transforms for interferograms calculated for the three optical responses of Fig. 1(c) are shown in Fig. 2(e). Next to the center branch at t=0 and the branch at t=26 fs, which were also observed in Fig. 1, two new branches appear at a longer delay time, which are related to interference with TR[18]. These have an asymmetrical response that reflects the instantaneous emission of TR. For both peaks, an exponential tail extends towards the lower time range, which reflects the fact that the resonant pulse arrives at an earlier time than the short pulse from the TR. In the spectral pulse interferometry analogy, these features are the cross correlate of the `sample pulse' from the nanoparticle, and the near-instantaneous TR pulse. Finally, we can use the broadband TR emission as a reference to reconstruct the phase of the particle emission[21,23,24]. In Fig. 2(c), we subtract the resonance behavior (Fig. 2(b)) from the nonresonant behavior (Fig. 2(a)) and see that the checkerboard pattern is shifted with respect to the nonresonant reference around the resonance wavelength. This is a result of the $\pi$ phase shift in resonant scattering across a Lorentzian spectral profile.

To experimentally determine what information CL interferometry reveals about the temporal characteristics of resonant CL excitation, we perform AR spectral CL measurements and compare CL interferograms for plasmonic and dielectric resonant scatterers. To study scattering from Au nanoparticles spaced a few microns above a reflecting surface, we fabricate a sample comprised of a 10-µm-diameter W wire on which 100-nm-diameter Au nanoparticles are coated. The particles support a LSPR at 550 nm wavelength.

We measure the CL interferogram in two configurations where the e-beam is placed either on the particle or grazing the particle at approximately 5 nm from its surface (see Supplementary Material (SM) for the CL spectra). When the e-beam grazes the particle, it additionally excites TR at the Au substrate surface, resulting in a richer interferogram. Figures 3(a) and 3(d) show the CL interferogram for the on-particle and off-particle geometries, respectively. We clearly see the expected interference, analogous to the calculated result in Fig. 1(a) and 2(a), strongly modulated by the dipolar emission of the Au particle.

We then take the Fourier transform of Fig. 3(a,d), as shown in Fig. 3(b,e). For the on-particle geometry, the interference is associated with the particle emission interfering with its reflection, which allows us to fit a distance between the particle and the substrate of $5.4\ \mu\mathrm{m}$. In the off-particle case, we observe additional peaks associated with the contribution of TR, as predicted in the analysis of Fig. 2. To further study the time characteristics of the Fourier peaks, we take line scans of the Fourier transform for the on-particle (Fig. 3(c)) and off-particle (Fig. 3(f)) CL signals at emission angles of $50°$ (orange) and $60°$ (purple). For the on-particle case, we focus on the peak at 30 fs (associated with the interference between the particle emission and its reflection), while for the off-particle case, we focus on the peak centered at approximately 70 fs (associated with the interference between the particle emission and TR). We clearly see that the on-particle scenario shows a broad peak in the Fourier domain. Furthermore, for the off-particle configuration, the peak is asymmetric, featuring a slower rise time than fall time.

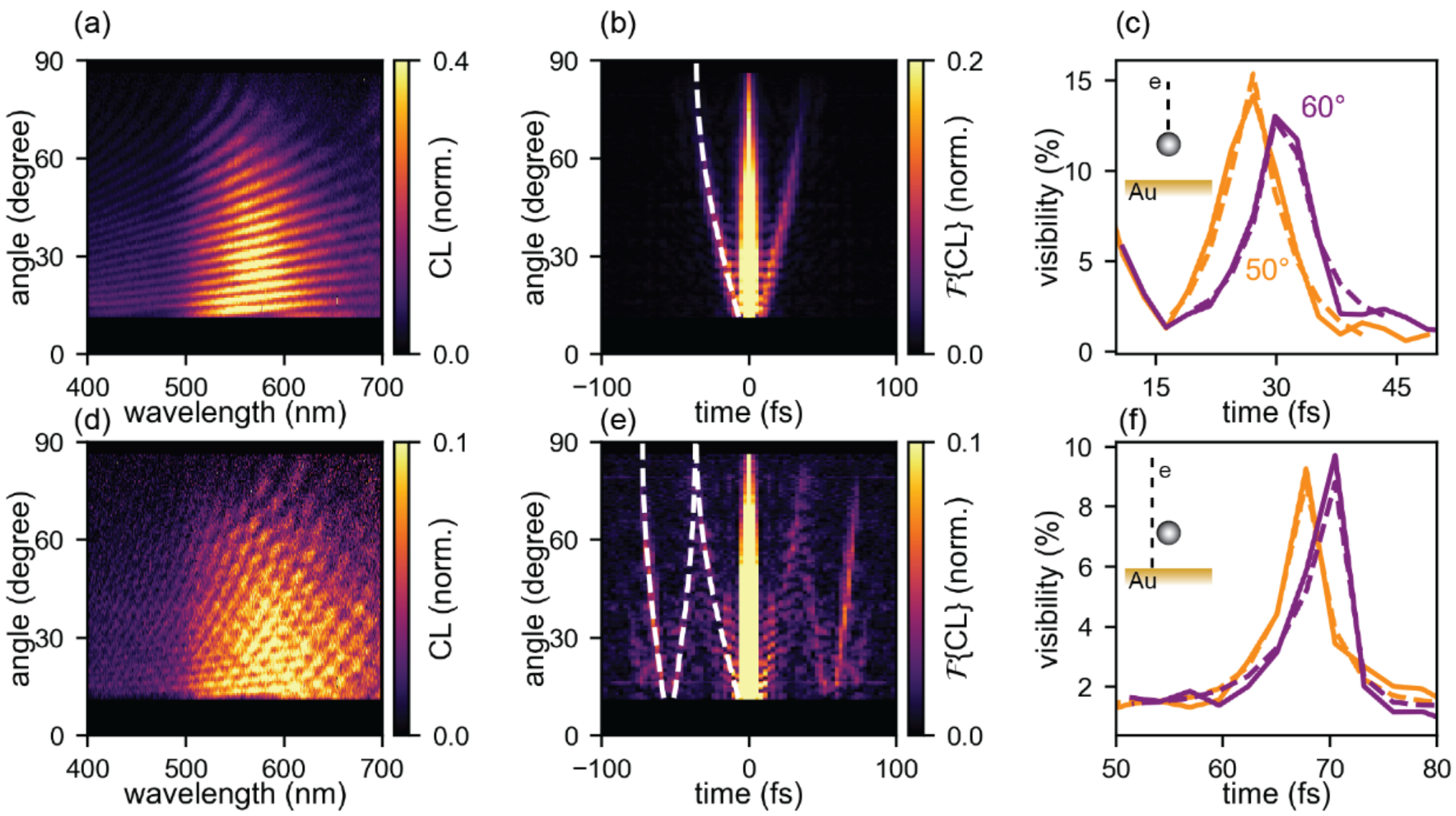

*Figure 3. Measured CL interferograms for a Au particle of 100 nm in diameter, placed above a Au surface. (a) Angle- and wavelength-resolved CL interferogram for on-particle excitation, (b) its Fourier transform to the time domain, and (c) line scans along emission angles of 50° (orange) and 60° (purple). (d-f) Same as (a-c), but off-particle excitation (see inset in (f)). The line scans in (c,f) are fitted to an exponential rise and fall time (dotted line), and the fitting parameters are shown in Table 1.*

To quantify this difference, we fit the peaks to the following function with a rise time $\tau_{\text{rise}}$ and fall time $\tau_{\text{fall}}$:

$$y(t) = \begin{cases} y_0 + Ae^{\frac{t-t_0}{\tau_{\text{rise}}}}, & \text{for } t \le t_0, \\ y_0 + Ae^{\frac{-(t-t_0)}{\tau_{\text{fall}}}} & \text{for } t \ge t_0, \end{cases} \tag{4}$$

where $t_0$ is the peak center, $y_0$ is the baseline intensity, and $A$ is the peak height. The fitting parameters are summarized in Table 1. The rise and fall times for the on-particle configuration are identical within the error margin. However, for the off-particle configuration, we see that, for the 60° measurement, the fall time is significantly lower than the rise time. This reflects the fact that the feature at hand is analogous to a pulse cross correlation of the instantaneous TR emission and the more slowly decaying emission from the Au particle. We note that the temporal resolution is limited by the spectral bandwidth of the spectrometer, which corresponds to 2.7 fs.

*Table 1. Fitting parameters of the data in Fig. 3 to an exponential rise and fall time according to Eq. (4).*

| Fitting parameter | Emission angle | $\tau_{\text{rise}}$ (fs) | $\tau_{\text{fall}}$ (fs) |
|---|---|---|---|
| On particle | 50° | 4.6±1.3 | 4.9±1.1 |
| | 60° | 3.9±1.1 | 3.8±1.3 |
| Off particle | 50° | 3.0±0.3 | 2.4±0.9 |
| | 60° | 4.0±0.8 | 1.3±1.8 |

To further investigate the effect of the resonant temporal response on the interferograms, we compare three samples with a different resonant optical response: 1) a plasmonic Au tip with a broad resonance, made by growing a Pt wire using e-beam induced deposition (EBID) and subsequently coating with 50 nm of Au; 2) chemically synthesized Au nanostars[25] placed on a W wire as described above, with a tip resonance at 700 nm and having a high quality factor; and 3) single-crystalline Si nanospheres[26] with multiple Mie resonances within the measured spectral range[27,28], also placed on a W wire (see SM for the CL spectra of the different samples and fabrication methods). In all cases, the e-beam is placed onto the structure, making sure that there is only a single source of excitation in the system, and we observe only the interference between the direct particle emission and its reflection (i.e., in the absence of TR).

We take the Fourier transform to study the behavior in the time domain and compare the three samples at two emission angles: $40°$ (orange) and $80°$ (purple). Figure 4 shows the Fourier transform of the experimental data (solid curves) for (a) the plasmonic tip, (b) the tip of a Au nanostar, and (c) the Si nanosphere. The heights of the nanoparticles above the substrate are 5.37 $\mu m$, 6.6 $\mu m$, and 9.2 $\mu m$, respectively. The temporal peaks shift to longer time delays for larger distances. Comparing the widths of the peaks, we see clear differences, with the sharpest peak observed for the plasmonic tip, and the broadest distribution for the Si nanosphere.

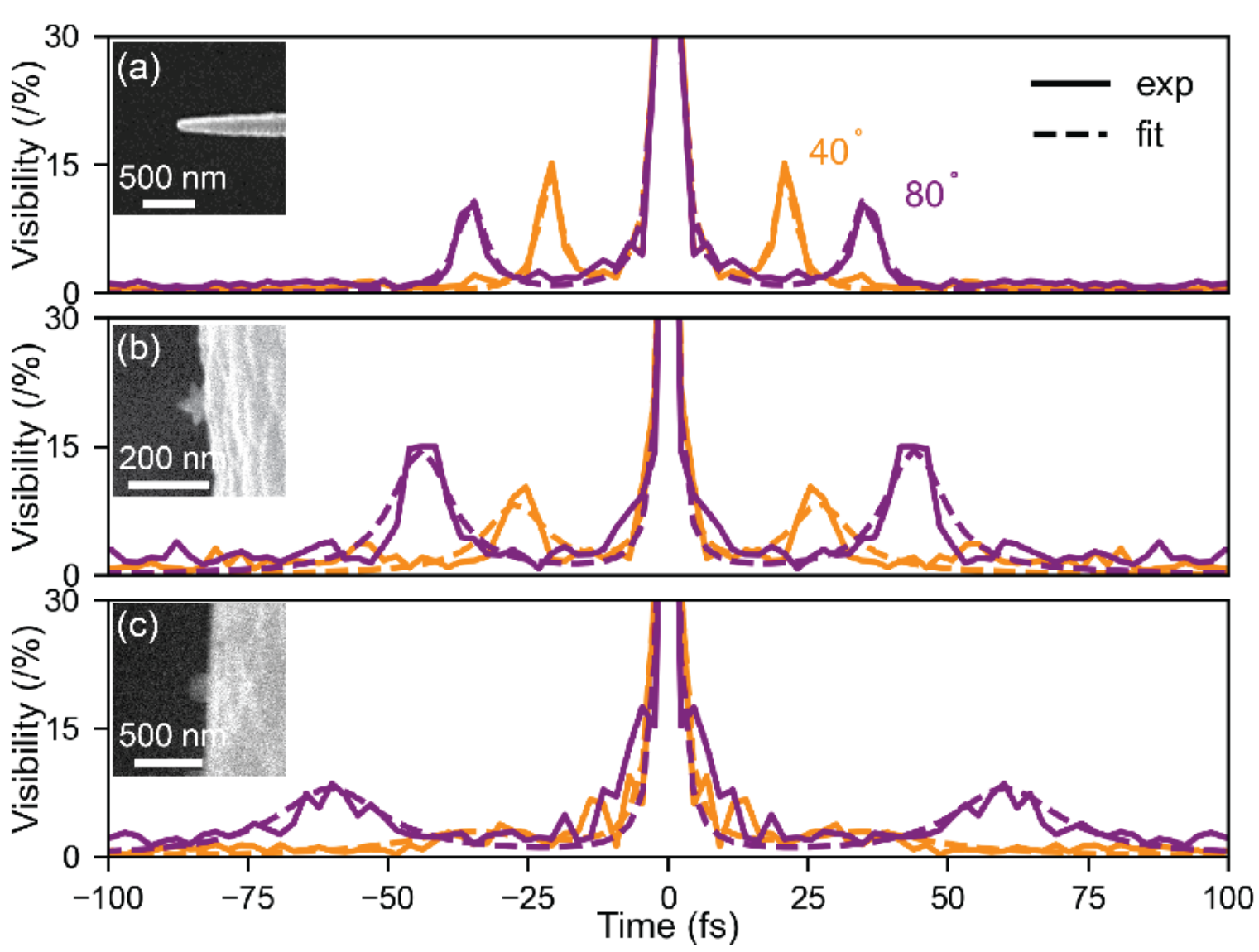

*Figure 4. Temporal distributions derived from CL interferograms for (a) a plasmonic tip, (b) a Au nanostar, and (c) a Si nanosphere. SEM images of the particles are shown in the insets. We plot experimental data (solid curves) along with fitted Lorentzian spectra (dashed curves) at two CL emission angles of, 40◦ (orange) and 80◦ (purple).*

In Fig. 4, we fit the spectra to three Lorentzian peaks (dashed curves). The fitting parameters are shown in Table 2. For the plasmonic tip, the peak full-width-at-half-maximum is 2.5 fs. This matches the broad spectral emission from the tip, resulting in a fast decay time of a TR-like emission of only a few optical cycles[29]. In contrast, for the Au nanostar, the temporal width is much larger, around 7 fs, corresponding to the spectrally sharp resonance at 700 nm. Finally, for the Si nanosphere, which supports multiple resonances in the spectral domain, we observe even broader peaks. For CL emission at an emission angle of $80°$, the fitted width is 9.8 fs, indicating a higher quality factor associated with the Mie resonances in the Si particle. The small differences in decay times observed for the two angles are within the temporal resolution of the measurements. Variations in coupling of near fields to the environment and varying angular distributions for specific modes could affect the decay time when changing the emission angle.

Interestingly, we discern oscillations within the individual peaks, associated with interference of the multipolar resonances in the spectrum of the Si particle, as predicted by the calculations shown in Fig. 1(d)[27,28,30]. The interval between two consecutive minima is approximately 7 fs, corresponding to a 0.6-eV energy splitting of the resonances, close to the 0.5 eV spectral splitting of the Si CL spectrum (see SM for the spectra).

*Table 2. Fitting parameters of the data from Fig. 4 to multiple Lorentzian peaks.*

| Fitting parameter | Emission angle | Au tip | Au nanostar | Si nanosphere |
|---|---|---|---|---|
| Width (fs) | 40° | 2.1±0.1 | 8.1±0.1 | - |
| | 80° | 2.9±0.1 | 6.1±0.2 | 9.8±0.5 |
| Visibility ($A/A_0$) (%) | 40° | 14 | 8 | - |
| | 80° | 11 | 14 | 8 % |

In conclusion, we have demonstrated that cathodoluminescence interferometry provides direct access to the temporal characteristics of resonant nanoscale emitters within the 1-10 fs time range. We showed analytically that the Fourier transform of CL interferograms encodes information on the temporal response and phase of optical resonances. Resonant emitters exhibit broadened temporal features that reflect their spectral linewidth, while multimodal systems give rise to characteristic temporal beating signatures associated with spectral mode splitting. By employing TR as a broadband reference, we further demonstrated cross-correlation measurements between instantaneous and resonant emission processes, resulting in asymmetric temporal responses. Experiments on plasmonic and dielectric scatterers validate the analytical model and establish a direct correspondence between spectral and temporal observables. These results establish CL interferometry as a versatile approach for probing the temporal response of scatterers with nanometer spatial resolution and provide a general framework for phase-sensitive studies of complex nanophotonic systems using CL spectroscopy.

## Supporting Information

See Supplementary Material for the CL spectrum of a Au nanoparticle excited in the center or with a grazing e-beam passing 5 nm away from the particle surface, as well as CL spectra for the Au nanotip, the tip of a Au nanostar, and the Si nanoparticle. We also provide additional information on the experimental methods.

## Conflicts of interest

The authors declare the following competing financial interests: Albert Polman is cofounder and co-owner of Delmic BV, a company that produces commercial cathodoluminescence systems like the one that was used in this work.

## Data availability statement

The data that supports the findings of this study are available from the corresponding author upon reasonable request.

## Funding Information

This work is financed by the Dutch Research Council (NWO) and has received funding from the European Research Council (ERC) under Grant Agreements Nos. 101019932 (QEWS) and 101141220 (QUEFES). This work was partly supported by JSPS KAKENHI Grant No. 24K01287 and Kobe University Strategic International Collaborative Research Grant.

– Supplementary Material –

# Revealing time characteristics of optical excitations in dielectric and plasmonic structures through cathodoluminescence interferometry

Evelijn Akerboom[a*], Hiroshi Sugimoto[b], Minoru Fujii[b], Nicolas Pazos-Perez[c], Ramon A. Álvarez Puebla[c,d], A. Femius Koenderink[a], F. Javier García de Abajo[d,e], and Albert Polman[a*]

[a]Center for Nanophotonics, NWO-Institute AMOLF, Science Park 104, 1098 XG Amsterdam, the Netherlands

[b] Department of Electrical and Electronic Engineering, Graduate School of Engineering, Kobe University, Rokkodai, Nada, Kobe 657-8501, Japan

[c] Department of Physical and Inorganic Chemistry, Universitat Rovira i Virgili, 43007, Tarragona, Spain

[d] ICREA-Institució Catalana de Recerca i Estudis Avançats, Passeig Lluís Companys 23, 08010 Barcelona, Spain

[e] ICFO-Institut de Ciencies Fotoniques, The Barcelona Institute of Science and Technology, 08860 Castelldefels, Barcelona, Spain

* Corresponding email: e.akerboom@amolf.nl, a.polman@amolf.nl

---

## S1 – CL spectra of a 100-nm-diameter Au particle

Figure S1 shows the cathodoluminescence (CL) spectra of a 100-nm-diameter Au nanoparticle excited by a 30-keV electron beam. Spectra were recorded for two excitation geometries: with the electron beam incident at the particle center (orange curve) and with the beam passing approximately 5 nm outside the particle surface (purple curve). The spectra correspond to the data presented in Fig. 3 of the main text and are averaged over all emission angles.

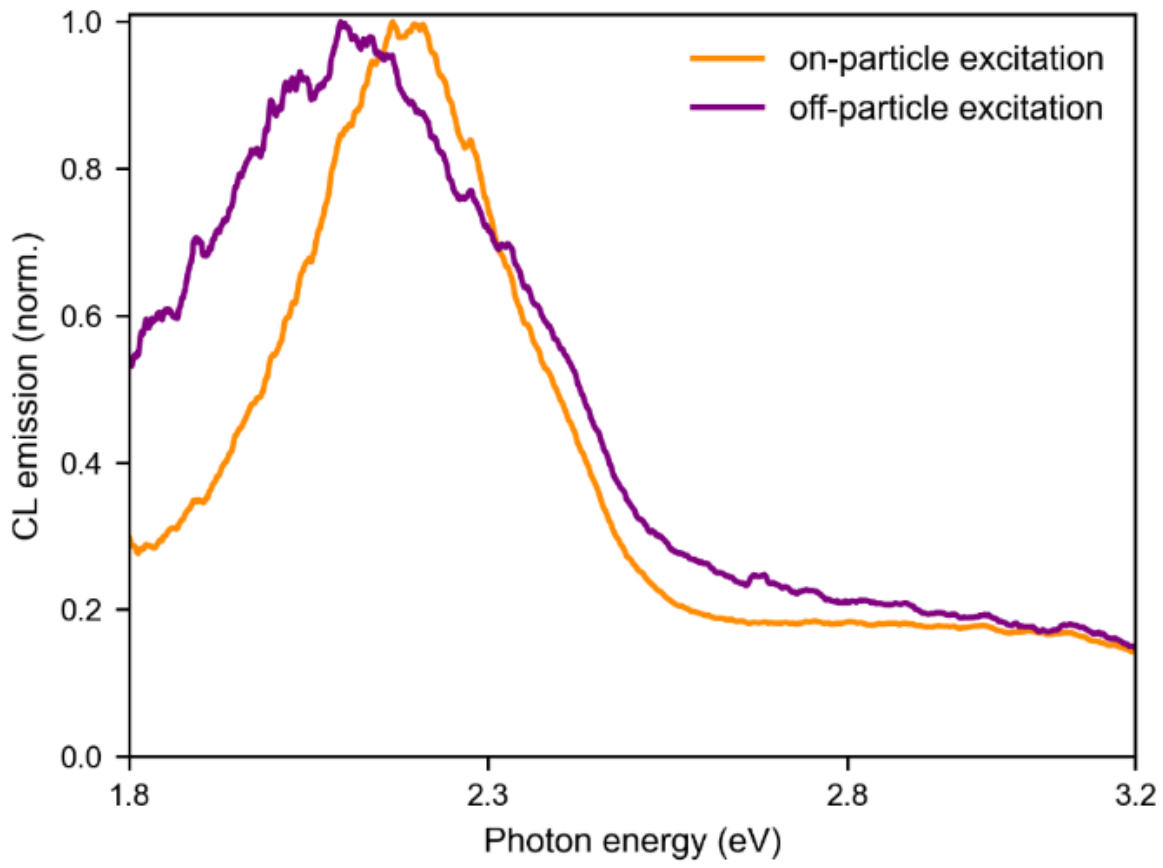

Figure S1. Normalized CL spectra of a 100-nm-diameter Au nanoparticle excited by a 30-keV electron beam. The spectra are shown for on-particle excitation at the particle center (orange curve) and off-particle (aloof) excitation with an impact parameter of approximately 5 nm relative to the particle surface (purple curve).

## S2 – CL spectra for all particles under consideration

Figure S2 shows the CL spectra of the three structures investigated in Fig. 4 of the main text: a lossy Au tip (orange curve), a tip in a Au nanostar (purple curve), and a Si nanosphere with a diameter of 190 nm (red curve). In all cases, the structures were excited by a 30-keV electron beam incident at the center of the structure. The spectra are averaged over all emission angles and correspond to the data presented in Fig. 4 of the main text.

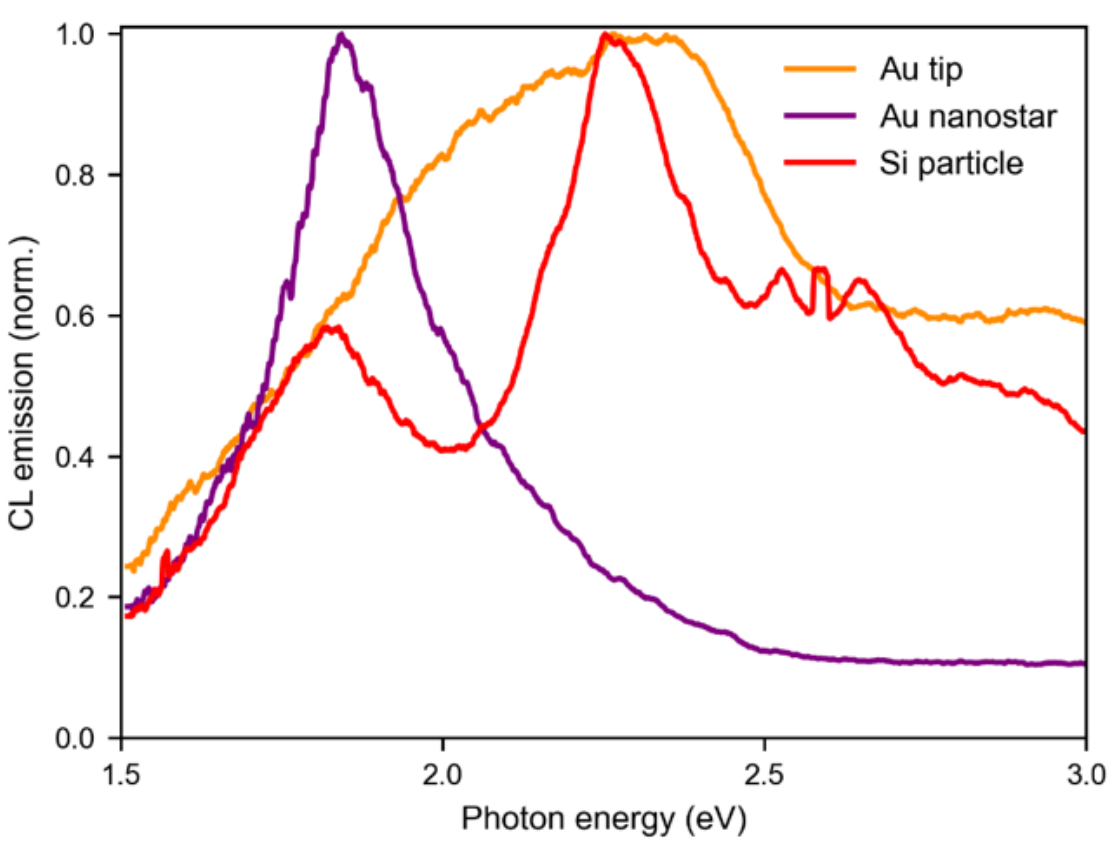

Figure S2. Normalized CL spectra obtained under on-particle excitation with a 30-keV electron beam for a lossy Au tip (orange curve), a tip in a Au nanostar (purple curve), and a 190-nm-diameter Si nanosphere (red curve).

# S3 – Experimental methods

## Sample preparation

Commercially available Au nanospheres were obtained from nanoComposix, San Diego, CA, USA.

The Au nanostars were chemically synthesized following a seed-mediated growth procedure. Au seed particles were first prepared using a modified Turkevich method, then stabilized with polyvinylpyrrolidone (PVP) and transferred to ethanol. Nanostar growth was initiated by adding the PVP-coated seeds to an N,N-dimethylformamide (DMF) solution containing $HAuCl_4$ and PVP. The formation of nanostars was indicated by a color change of the solution from red to blue, consistent with the development of a red-shifted plasmonic response. The reaction mixture was stirred overnight to ensure complete reduction of the gold precursor. The optical response of the nanostars was controlled by adjusting the relative concentrations of the seeds, $HAuCl_4$, and PVP.[1]

The Si nanospheres were prepared from silicon powder obtained by mechanically crushing SiO feedstock. The powder was annealed in a nitrogen atmosphere and subsequently etched using hydrofluoric acid (HF). Individual Si nanospheres were suspended in methanol, ultrasonicated, and size-selected by filtration, resulting in particles with diameters of approximately 200 nm.[2]

## Fabrication of elevated particles

To create elevated nanoparticle configurations, a circular tungsten (W) wire with a diameter of 10 µm was used as a spacer. The nanoparticles were dispersed in isopropanol (IPA) and ultrasonicated to minimize agglomeration. The W wire was cleaned in IPA and dried under a nitrogen flow before being immersed in a droplet of the nanoparticle suspension deposited on a glass substrate. After complete solvent evaporation, the particle-coated wire was transferred onto a Au-coated Si substrate. A 10-µL droplet of deionized water was used to improve adhesion of the wire to the substrate. The sample was subsequently dried at 60 °C. Owing to the curvature of the wire, the nanoparticles were positioned at heights of approximately 5 µm above the Au surface, with the exact height depending on the particle location along the wire circumference.

## Fabrication of free-standing Au tip

Single-crystal polished Si(100) substrates were cleaned using a base-piranha procedure and subsequently coated with a 2-nm Cr adhesion layer and a 50-nm Au film by sputter deposition (EM ACE600, Leica Microsystems). Free-standing Pt nanopillars were fabricated by electron-beam-induced deposition (EBID) in a scanning electron microscope (Helios NanoLab 600, FEI/Thermo Fisher Scientific). The deposition was performed using a 2-keV electron beam with a current of 170 pA, a dwell time of 1.5 ms, and a stage tilt angle of 45°. The resulting pillars had a base diameter of approximately 300 nm that tapered to a tip diameter of about 40 nm. Following EBID, the entire sample was coated with an additional 50-nm Au layer by sputter deposition, resulting in free-standing Au nanotips positioned above a 100-nm thick Au film.

## Angle-resolved spectral CL measurements

The angle-resolved (AR) spectral measurements were performed in a FEI Quanta FEG 650 scanning electron microscope (Thermo Fisher Scientific Inc., MA, USA), equipped with a parabolic mirror to collect the AR spectral CL. Using a vertical slit, only a narrow radial angle was selected and projected onto a spectrometer to acquire the CL data with both angular and spectral information simultaneously. For the AR spectral CL measurements, the electron energy was 30 keV, using a current of 1.4 nA. The AR spectral CL interferograms were measured using an acquisition time of 240 s for Au nanoparticles, Au nanostars, and Si nanoparticles, and 360 s for the Au nanopillar. For all data, a dark measurement representing background radiation was subtracted, and the spectrum was corrected for the system's response using TR emission from a single-crystal Al sample as a reference.